\documentclass[aps,amsfonts,nofootinbib,twocolumn]{revtex4}
\usepackage{amsmath}
\begin{document}
\title{
Modified Special Relativity on a fluctuating spacetime
}
\author{R.  Aloisio}
\affiliation{INFN - Laboratori Nazionali
del  Gran  Sasso,  SS.  17bis,  67010  Assergi  (L'Aquila)  -  Italy}
\author{A.  Galante}
\affiliation{INFN - Laboratori Nazionali
del  Gran  Sasso,  SS.  17bis,  67010  Assergi  (L'Aquila)  -  Italy}
\author{A.F.  Grillo}
\affiliation{INFN - Laboratori Nazionali
del  Gran  Sasso,  SS.  17bis,  67010  Assergi  (L'Aquila)  -  Italy}
\author{S. Liberati}
\affiliation{International School for Advanced Studies and INFN, Via
Beirut 2-4, 34014, Trieste - Italy}
\author{E.  Luzio}
\affiliation{Dipartimento di Fisica,  Universit\`a di L'Aquila, Via
Vetoio  67100 Coppito (L'Aquila) - Italy}
\author{F.  M\'endez}
\affiliation{ Departamento de Fisica, Universidad de Santiago de
  Chile, Av. Ecuador 3493, Casilla 307 Stgo-2 - Chile}

\begin{abstract}
It  was  recently  proposed  that  deformations  of  the  relativistic
symmetry, as  those considered  in Deformed Special  Relativity (DSR),
can be seen as the outcome  of a measurement theory in the presence of
non-negligible      (albeit      small)     quantum      gravitational
fluctuations~\cite{libson,libnoi}.   In  this   paper   we  explicitly
consider the  case of a spacetime  described by a  flat metric endowed
with stochastic  fluctuations and, for  a free particle, we  show that
DSR-like nonlinear  relations between the  spaces of the  measured and
classical  momenta, can  result  from the  average  of the  stochastic
fluctuations  over a scale  set be  the de  Broglie wavelength  of the
particle.  As   illustrative  examples  we   consider  explicitly  the
averaging procedure  for some simple stochastic  processes and discuss
the physical implications of our results.

\end{abstract}
\maketitle
\section{Introduction}
The existence  of a length (or  energy) scale at which  the effects of
quantum  gravity (QG) can  not be  neglected, and  the fact  that this
scale should be the same for all the inertial observers, have recently
led  to the  proposal  of a  Deformed  (or Doubly)  theory of  Special
Relativity (DSR) \cite{dsr,ms,alu}.  In momentum space a simple way to
characterize DSR theories is  that they corresponds to deformations of
the Poincar\`e algebra in the boost sector. Indeed such deformed boost
algebra amounts to the assertion  that physical energy and momentum of
DSR can  be always  expressed as nonlinear  functions of  a fictitious
pseudo-momentum $\pi$,  whose components transform  linearly under the
action  of  the Lorentz  group~\cite{ms,JV}.  More  precisely one  can
assume  the  existence of  an  invertible  map  $\cal F$  between  two
momentum spaces: the {\em  classical space} $\cal P$, with coordinates
$\pi_\mu$, where the Lorentz group  acts linearly and the {\em physical
  space} $P$,  with coordinates $p_\mu$, where the  Lorentz group acts
as the  image of its action on  ${\cal P}$.  Also, ${\cal  F}$ must be
such  that  ${\cal  F}:[\pi_0,\vec{\pi}]\rightarrow  \kappa$  for  all
elements   on   ${\cal   P}$   with   $|\vec   {\pi}|=\infty$   and/or
$\pi_0=\infty$    where    $\kappa$    is   some    quantum    gravity
scale~\footnote{Note that  this requirement  imposed on the  theory is
  equivalent to say  that such a scale should be the  same for all the
  inertial observers.}.  Most commonly such a scale is taken to be the
Planck   energy,   $\kappa   \sim   M_{\rm   P}   \approx   1.22\times
10^{19}\;$GeV.

While  the precise  formulation  in  the momentum  space  for the  one
particle case seems  very well understood \cite{ms,alu,dsrmom}, there
are still pressing  open problems that makes unclear  the precise role
of DSR  in QG \cite{dsrqg}.   In particular it  is so far  unknown the
realization of  DSR in coordinate  space, (although great  efforts has
been devoted  to find a theory  mirroring what happen  in the momentum
space~\cite{ms,sptm})  and even  the  theory in  momentum space  faces
severe  challenges as  the problem  of saturation  (the fact  that DSR
apparently does  not allow for  objects with energies grater  than the
invariant scale, the Planck  momentum) and the  multiplicity problem
(the  fact  that   in  principle  there  is  an   infinite  number  of
deformations of relativistic symmetry).

These  problems appear  to be  related  to the  interpretation of  the
non-linear    relation   between    the    classical   and    physical
momenta~\footnote{An   interpretation  that  indeed   should  actually
  justify why the physically relevant variables $p$ are indeed the DSR
  ones, and  not the much more familiar,  {\em classical} momenta
  $\pi$.} and have recently motivated the development of approaches to
DSR  within which  the latter  is seen  no more  as a  new fundamental
extension  of   special  relativity   but  instead  as   an  effective
modification of  it due to  the non complete negligibility  of quantum
gravitational  effects  as the  energy  scales  involved approach  the
quantum   gravity  one.   For   example  one   of  the   most  pursued
interpretations  of   DSR  actually  resort  to   a  more  fundamental
five-dimensional momentum  space in  order to make  sense of  the four
dimensional   non-linear   dispersion   relations  that   characterize
DSR~\cite{5D}.

Along this line of  though, it was recently proposed~\cite{libson} that
DSR could be interpreted  as an effective theory of measurement for
high  energy particle  properties.  According  to this  framework, the
relation between ``true"~ energy and  momentum of a particle (the {\em
  classical} variables $\pi$ of DSR) and observed quantities (the {\em
  physical}  variables  $p$ of  DSR)  acquires,  at sufficiently  high
energies,  Planck suppressed  distortions induced  by  quantum gravity (QG)
effects.    These   relations   can   be  identified   with   DSR-type
deformations.  In \cite{libnoi} we  argued that this non linear nature
arises as a  result of the unavoidable averaging  over QG fluctuations
of  the metric around  flat spacetime  which is  required in  order to
properly define energy  and momentum in first place.   In this note we
go a step further and propose  an explicit toy model of how this class
of non linear functions might appear.

Our model is based on three points
\begin{itemize}
\item we take into account  quantum gravity fluctuations at the Planck
  scale by  adding a  stochastic term to  the classical  1-form tetrad
  field (associated to 
  the  spacetime  metric in  the  system  of  reference in  which  the
  measurement is performed~\cite{libson,libnoi}). 
\item we  define an average over the  spacetime fluctuations performed
  by  the particle  as it  propagates as  the result  of  its (quantum
  mechanical) finite size,
\item we make an assumption about the specific form of the correlation
  of these stochastic process.
\end{itemize}

The first  point is,  we believe,  a quite reasonable  one, and  it is
introduced in order to model  the spacetime fluctuations at the Planck
scale. Although some arbitrariness is present in the assumption of the
kind of modification  of the tetrad, our result  will be quite generic
and one might  hope that the precise form of  such a modification will
be  in the  end  fixed by  some  quantum gravity  theory.  The  second
assumption is at  the heart of our model and it  asserts that what the
particle  really  does  as  it  propagates  is  to  average  over  the
fluctuations, defined with respect to the system of reference in which
the particle is observed, and such averaging is just determined by the
particle  characteristic ``resolution''  scale which  we assume  to be
related to  its de Broglie wavelength.   The third point  is, at least
partly, a consequence of the first one, as we will see, but clearly is
also due to our ignorance on  the details of quantum gravity. In other
words, we have  to choose some specific stochastic  process.  

This toy model is clearly a  poor substitute for a full QG theory, but
it  will  allow  us to  elaborate  on  the  possible outcomes  of  the
procedure and  have an explicit example  of how things  might work. In
the next section we will  introduce the fluctuations in a very general
ground.  In section  III we will provide examples  of fluctuations and
its relations  with DSR. The  final section is devoted  to discussion,
some speculative issues and conclusions.

\section{Fluctuations and DSR}
Our underlying assumption is that the spacetime, when probed at scales
of the order  of the Planck one, reveals its  quantum nature which can
be modeled  by a classical metric with
an  extra   fluctuating  part  \cite{fluct}.  Let   us  consider  four
dimensional spacetimes 
with topology  $\mathbb{R}\otimes {\cal M}_3$, where  ${\cal M}_3$ are
space-like three manifolds parametrized by a real number $t$, which we
will call  time. The  usual ADM decomposition  \cite{adm} of  the four
dimensional metric  $G_{\mu\nu}$ (with $\{\mu,\nu\}\in \{0,1,...,3\})$
reads
\begin{equation}
\label{adm}
G_{\mu\nu}=\left(
\begin{array}{cc}
-N^2+N^lN_l & N_m \\
N_n & g_{mn}
\end{array} \right),
\end{equation}
where $g_{mn}$ is the metric of ${\cal M}_3$ and $N^lN_l=N_lN_mg^{lm}$
(with  $g^{mn}$  being  the  inverse  of $g$  and  the  latin  indexes
$\{m,n,\dots\}\in \{1,2,3\}$  being indexes of  coordinates adapted to
$M_3$).  The lapse ($N$)  and shift ($N_m$) functions characterize the
embedding of $M_3$  in the four dimensional spacetime.

In the Hamiltonian formulation of  gravity, the metric $g$, as well as
its  conjugate momenta,  are determined  by the  equations  of motion,
while the  lapse and  shifts remain arbitrary  (Lagrange multipliers).
An alternative formulation of gravity  --- which is the starting point
for Loop  Quantum Gravity --- can  be given in terms  of the one-forms
$e^A=e^A\,_\mu\,dx^\mu$,           where          the          indexes
$\{A,B,\dots\}\in\{0,1,2,3\}$ are local  Lorentz indexes, that is, the
tetrads are defined up a local Lorentz transformation $\Lambda^A\,_B$:
$e'^{A}=\Lambda^A\,_B\,e^B$  and  characterize   a  given  system  of
reference.

The  matrix  element $e^A\,_\mu$,  for  the  previously mentioned  ADM
decomposition (\ref{adm}) is
\begin{equation}
\label{teadm}
e^A\,_\mu=\left(
\begin{array}{cc}
e^0\,_0 &0 \\
e^L\,_0 & e^L\,_m
\end{array}\right),
\end{equation}
where  $\{I,J\dots\}\in\{1,2,3\}$ and  the  following relations  hold:
$e^0\,_0=N,\,\,\,N_m=e^I\,_m                  e^J\,_0\,\eta_{IJ},\,\,\,
g_{mn}=e^I\,_m\,e^J\,_n\,\eta_{IJ}.$  Here  $\eta_{IJ}$=diag$(1,1,1)$.
Note also  that the requirement $  {e^0}_i=0$ is equivalent  to say that
the  ``time''   coordinate  in  spacetime  coincides   with  the  time
coordinate in the tangent manifold.  Finally  $e^\mu\,_A$ are
the  matrix elements  of the  inverse of  (\ref{teadm}), that  is, the
solution  of  $e^A\,_\mu   \,e^\mu\,_B=\delta^A\,_B$.   {  Within  the
framework proposed  in \cite{libson,libnoi} the  observed momenta will
then  be identified as  $p_{A}=\pi_{\mu}\langle {e^{\mu}}_{A}  \rangle $
where the  $\langle \dots \rangle$  implies the averaging over  the QG
fluctuations.}

In the  example we study  here, we will  assume that spacetime  has an
underlying structure due to  quantum effects.  A ``free" particle that
moves  in such  a spacetime  probes this  structure  ({\em e.g.}~moves
along  slightly deformed  geodesics)  with a  resolution  that can  be
assumed  to be  related (and approximately equal) 
to the de  Broglie wave  length  of the
particle, since this is the best accouracy with which the latter can
be localized~\footnote{  It  is  possible,  however,  to  conceive
different choices of the  resolution with which the particle ``probes"
spacetime which are corresponding  to decreased accouracy. For example
one can imagine that the spacetime structure is probed by the particle
via gravitational interaction, in this  case the natural scale come up
to  be  the curvature  radius  associated  with  the particle  $M_{\rm
P}/{\cal  E}^2$ which  corresponds  to  a loss  of  accuracy of  order
$M_{\rm P}/{\cal E}$  with respect to the de  Broglie wavelength. This
seems to be the preferred  case in some alternative derivations of the
modified dispersion relations of DSR~\cite{GLPR}.}.  In this sense the
particle   in  its   motion  averages   the  fluctuations   over  this
length scale.   In agreement with the approach of \cite{libson,libnoi}  
the outcome  of  such averaging  has to  be
calculated from the  form of such fluctuations in  the reference frame
in  which the  particle energy  and momentum  are measured~\footnote{A
frame dependence would seem to  appear in this statement. This will be
further discussed later.}.

In  modeling the above framework  we shall have  to make assumptions
about the  form of  the quantum fluctuations.  In this sense  we shall
here assume that the metric can be described as the sum of a classical
term plus a stochastic one whose role is to model our ignorance of the
details  of the  underlying  quantum gravity  theory.  Within our  ADM
approach  the above  ansatz  implies  that the  tetrad  field will  be
determined up  to a function which  is a stochastic  function of time:
$\xi(t)$  (indexes are not  important at  this point).  Since it  is a
stochastic  function  there  is  a probability  distribution  $P[\xi]$
associated to each one of these processes.

Form  the operational  point  of  view, this  stochastic  term can  be
understood as follow. Let assume that an observer is able to determine
the tetrad at some instant  $t_1$. If the same experiment is performed
at  $t_2>t_1$,  then  there  are  no  ways to  predict  the  value  of
$\xi(t_2)$  from our  knowledge  of $\xi(t_1)$.  The only  information
available are  the correlation  of such values.  Moreover if  we could
have a  large number of copies  of the same  experiment ({\em i.e.}~an
ensemble), then  all the  so obtained tetrads  would differ by  a term
which distributes according to $P[\xi]$.

Since  we are  interested  in deformations  of  special relativity  we
restrict to the case  of flat classical space-like hypersurfaces, {\em
i.e.}~in the  absence of the  above mentioned stochastic  processes we
recover classical flat spacetime.  In this sense it is then reasonable
to cast the matrix elements $e^A\,_\mu$ in the form
\begin{equation}
\label{testoc}
e^A\,_\mu=\left(
\begin{array}{cc}
1+\xi^0\,_0(t) &0 \\
\xi^L\,_0(t) & \delta^L\,_m+\xi^L\,_m(t)
\end{array}\right),
\end{equation}
where $\xi^A\,_\mu  (t)$ are time dependent  stochastic processes.  In
what follows we will {\it  assume} that such stochastic processes have
a Gaussian joint probability distribution $P[\xi^A\,_\mu,t]$ with zero
mean value.   Of course  there is a  priori no physical  reason, apart
from  simplicity  and  computability,   for  the  choice  of  Gaussian
fluctuations.

In    order    to   preserve    rotation    invariance,   we    choose
$\xi^L\,_m=\delta^{L}\,_{m}\eta(t)$, with $\eta(t)$ is some stochastic
function.  Then the  matrix  elements of  the  tetrad with  stochastic
components reads 
\begin{equation}
\label{testocf}
e^A\,_\mu=\left(
\begin{array}{cc}
1+\xi(t) &0 \\
\xi^L\,_0(t) & {\bf 1}[1+\eta(t)]
\end{array}\right),
\end{equation}
where   ${\bf  1}=$diag$(1,1,1)$   and   $\xi(t)\equiv  \xi^0\,_0(t)$.
Eq.~(\ref{testocf}) can now be  interpreted as the standard tetrad for
flat spacetime plus time dependent corrections.

We now need  to introduce in our model the  finite resolution scale of
the  elementary particle probing  the spacetime.   The natural  way to
incorporate this information  is to sum up all  possible values of the
tetrad within  this resolution scale.  Generically, if  $y(t)$ is some
quantity  such that  we do  not have  information below  certain scale
$\Delta$, then we  define its {\it mean value} as  its average on this
resolution scale
\begin{equation}
\label{ave}
y(\Delta)=\frac{1}{\Delta}\int_{t_0}^{t_0+ \Delta} y(t)dt.
\end{equation}

We define, therefore, the {\it mean } tetrad $\epsilon^A\,_\mu (\Delta)$ as the
time average of the matrix elements of the tetrad (\ref{testocf}) in a given
resolution interval $\Delta$
\begin{equation}
\label{titet}
\epsilon^A\,_\mu(\Delta)=\left(
\begin{array}{cc}
1+\xi(\Delta) &0 \\
\xi^L\,_0(\Delta) & {\bf 1}[1+\eta(\Delta)]
\end{array}\right),
\end{equation}
where all quantities $\xi(\Delta)$, $\xi^L\,_0 (\Delta)$ and $\eta(\Delta)$
are defined by (\ref{ave}). 

The tetrad, therefore, has been averaged in order to take into account
the  finite  resolution, but  not  over  the  values that  the  random
variables can take --- actually $(-\infty,\infty)$ for every fixed
value of  $t$ --- hence the  inverse tetrad in principle  could not be
defined.   For  instance, the  probability  for  $\eta$  to be  on  an
interval containing the value $-1$ is  not zero --- although it can be
made as  small as  desired {\em e.g.}   in the  Gaussian approximation
above --- and  therefore, the probability for the  spatial part of the
tetrad of being  zero is finite which implies that  the inverse of the
tetrad is ill defined.

In order to circumvent this problem,  we define an ``effective
tetrad" as a tetrad that  is compatible with the metric obtained after
the ensemble average is performed. In fact, since in general $\langle
f(x)\,g(y) \rangle  \neq \langle f(x)  \rangle \langle g(y)  \rangle $
for a  given probability distribution  $P[x,y]$, it is clear  that the
effective  tetrad will not  be the  same as  the tetrad  obtained just
taking the ensemble average on (\ref{titet}).

 The  origin of  this problem,  from a  strictly statistical  point of
 view, resides on  the fact that we have  chosen fluctuations that can
 have any value,  even large, although with a  small probability. This
 can  be  formally  thought  to  mimic the  contribution  of  topology
 changing metrics to  the Path Integral of QG. The  choice of using an
 effective  tetrad  corresponds  in  a  sense  to  keeping,  of  these
 contributions, only the (relatively) large distance effects.

Therefore,  consider  the  metric $\langle  G_{\mu\nu}(\Delta)\rangle$
  obtained    from   the   ensemble    average   (denoted    here   by
  $\langle\cdot\rangle$)  of the quadratic  form constructed  with the
  mean  tetrad  (\ref{titet}).  The  effective  tetrad matrix  element
  $\bar{e}^A\,_\mu$ is defined through the relation
\begin{eqnarray}
\label{Gm}
\langle   G_{\mu\nu}(\Delta)\rangle  &  =&   \langle  {\epsilon^A}_\mu
(\Delta) ~{\epsilon^B}_\mu 
(\Delta) \rangle~\eta_{AB}, \nonumber
\\
\label{tetef}
&\equiv&  {\bar{e}^A}\,_\mu  ~  {\bar{e}^B}\,_\mu ~  \eta_{AB}.
\end{eqnarray}

{In  order to evaluate  this quantities explicitly, let  us assume
  that the stochastic variables $\xi^L\,_0(t)$, $\eta(t)$ and $\xi(t)$
  are   not   correlated.    That   means  that   mixed   terms   like
  $\langle\eta(\Delta)~\xi(\Delta)\rangle$                           or
  $\langle\xi(\Delta)~\xi^L\,_0(\Delta)\rangle$ 
  can not appear  in the final expression; also,  since we have chosen
  all processes of Gaussian type 
with mean value zero, linear  terms can not appear neither. A direct
  calculation shows that the effective metric turn out to be}
\begin{eqnarray}
\label{admmen}
&&\langle G_{\mu\nu} (\Delta) \rangle =\nonumber
\\
&=&\left(
\begin{array}{cc}
 \langle \vec{\xi}(\Delta) \cdot \vec{\xi}(\Delta) - [1+\xi(\Delta)]^2
 \rangle & 0\\
 0 & \langle [1+\eta(\Delta) ]^2 \rangle \delta_{ij}
\end{array} \right), \nonumber
\\
&=&\left(
\begin{array}{cc}
-1+ \langle \vec{\xi}(\Delta) \cdot \vec{\xi}(\Delta) -\xi^2(\Delta)
\rangle & 0\\
0 & [1+ \langle \eta^2(\Delta) \rangle ]\delta_{ij}
\end{array} \right), \nonumber
\\
&&
\end{eqnarray}
with $\vec{\xi}\cdot\vec{\xi}=\xi^L\,_0\xi^M\,_{0}\,\eta_{LM} $. 

It is clear that modifications to the metric come from the correlations of the
stochastic functions which have the general shape 
\begin{equation}
\label{medcorr}
\langle\alpha(\Delta)\beta(\Delta)\rangle=\frac{1}{\Delta^2}\int_{t_0}^{t_0
  + 
\Delta} \langle \alpha(s)\beta(s') 
\rangle dsds', 
\end{equation}
for $\alpha,\beta$ two generic stochastic process.
To avoid a dependence on $t_0$ we demand
translational  invariance on  the correlators.

Notice, moreover, that the effective  metric, after averages,  depends on  the
resolution scale of the particle,   much like the {\it rainbow gravity
models}  \cite{rainbow}.  This  dependence  appears only  through  the
correlation functions of the stochastic processes.  

{From} the effective metric it is possible to read the effective tetrad
and also its inverse.  It is not hard to see that they are given by
\begin{eqnarray}
\label{teteff}
 \bar{ e}^A\,_\mu &=&
\left(
\begin{array}{cc}
\sqrt{1+ \langle\xi^2 - \vec{\xi}\cdot\vec{\xi}\rangle_c }& 0\\
0 & \delta_{ij}\sqrt{1+ \langle\eta^2\rangle_c}
\end{array}\right),\nonumber
\\
&&
\\
 \bar{e}^\mu\,_A&=&
\left(
\begin{array}{cc}
\frac{\textstyle 1}  {\textstyle\sqrt{ 1+  \langle\xi^2  - \vec{\xi}  \cdot
    \vec{\xi} \rangle_c}} & 0\\
0 & \delta_{ij}\frac{\textstyle 1}{\textstyle \sqrt{1+ \langle \eta^2\rangle_c}}
\end{array}\right),\nonumber
\\
\label{inteteff}
\end{eqnarray}
where we have defined for every stochastic process the correlator
as (\ref{medcorr})
$$
 \langle\xi^2\rangle_c \equiv \frac{1}{\Delta^2}\int_{t_0}^{t_0       +
  \Delta} \langle \xi(s)\xi(s')\rangle \, dsds'. 
$$

To make contact with DSR interpretation of \cite{libson}, consider the
measured momentum of the particle $P_A$.  In our model, it is given by
the average on the previously defined stochastic processes, that is by
$ \pi_\mu\bar{ e}^\mu\,_A$. Therefore, the
measured energy $E$ and momentum ${\bf p}$ for a particle are given by
\begin{eqnarray}
E&=&\frac{{\cal   E}}{\sqrt{1+  \langle  \xi^2-\vec{\xi}\cdot\vec{\xi}
    \rangle _c}}  , \nonumber  \\ &\sim&{\cal E}  \left( 1-\frac{1}{2}
\langle \xi^2-\vec{\xi}\cdot\vec{\xi} \rangle _c + \cdots \right),
\label{meden}
\\  {\bf p}&=&\frac{\boldsymbol  \pi}{\sqrt{1+ \langle  \eta^2 \rangle
    _c}},\nonumber  \\   &\sim&{\boldsymbol  \pi}  \left(1-\frac{1}{2}
\langle \eta^2 \rangle _c+ \cdots \right).
\label{medmom}
\end{eqnarray}
The first line of both equations is, within our assumptions, in
principle valid to all orders in the correlators.
In passing from the first to the second line of both equations we have
restricted the corrective  terms to
satisfy    $    \langle    \eta^2    \rangle    _c    \ll1,    \langle
\xi^2-\vec{\xi}\cdot\vec{\xi}  \rangle _c\ll1$,  and it  is  useful to
remind here  that the stochastic  variables $\eta$ and  $\xi$ contain
the resolution  scale. Lacking a quantum mechanics of particles
in a QG background,  it is possible that the identification of the 
resolution scale with
the wavelength of the particle, namely its inverse energy, can only be
made at first order \footnote{It is however tempting to take
  Eq.~(\ref{meden}), (\ref{medmom}) literally to all orders, $ \Delta \propto
1/{\cal E} $ strictly and ask what are the conditions to obtain the known DSR
theories. For instance DSR2 would require the correlators to be equal and
diverging as ${\cal E}^2/\kappa^2 $ as $\cal E $ goes to infinity.}.

 Now  let us compare  with DSR cases.   DSR1 is
characterized  by  the  nonlinear  relation  (at first  order  in  the
invariant energy-momentum scale
$\kappa$)
\begin{eqnarray}
\label{dsr1fir}
E&=&{\cal E}\left(1-\frac{ {\cal E}}{2\kappa}\right) ,
\\
{\bf p}&=&{\boldsymbol \pi}\left(1-\frac{\cal E}{\kappa}\right),
\end{eqnarray} 
where we have  discarded the contribution proportional to
$m/\kappa$ in the first equation.

DSR 2, instead,  is characterized by the nonlinear
relations (also at first order in $\kappa$)
\begin{eqnarray}
\label{dsr2}
\label{dsr2e}
E&=&{\cal E}\left(1-\frac{\cal E}{\kappa}\right),
\\
\label{dsr2p}
{\bf p}&=&{\boldsymbol \pi}\left(1-\frac{\cal E}{\kappa}\right).
\end{eqnarray}

Comparing  expressions  (\ref{meden})   and  (\ref{medmom})  with  the
corresponding  ones   coming  from  DSRs,  we   see  that  appropriate
assumptions  about the  shape of  the correlators,  together  with the
identification  of the  resolution  scale with the inverse  classical
energy or  momentum (or more  in general with a  function of them
  and the Planck  scale), allows to obtain them  from this stochastic
model.  

Now  that we have  established the  connection between  our stochastic
model and DSR1 and  DSR2, we can turn to the problem  of which kind of
fluctuations gives  rise to specific  DSRs. In particular, we  need to
know how  the invariant  scale emerges. We  shall now deal  with these
issues by discussing explicitly some examples.

\section{Examples of Stochastic Process}

Since we have introduced stochastic processes at the level of tetrads,
modifications to the metric  come only from auto correlation functions
which  are  also responsible  of  the  specific  non linear  functions
appearing in (\ref{meden}) and (\ref{medmom}).
In other words,  to make contact with DSR we only  need to specify the
value of the  correlation, which, generically, is not enough  to completely determine
the stochastic process.

\subsection{Ornstein--Uhlenbeck process}
As  a  starting point,  let  us  consider  the Ornstein-Uhlenbeck
  process,  $\sigma(t)$. This  is  the only  Gaussian, stationary  and
  Markovian  process  and  has  a  correlation function  of  the  form
  \cite{stobook}
\begin{equation}
\label{oruh}
 \langle \sigma(s)\sigma(s') \rangle =\frac{D}{2k}e^{-k|s-s'|},
\end{equation}
where $D$ and $k$ are  dimensional constant. It is clear that $k^{-1}$
has dimensions of time, {\it i.e.}~the correlation time while $D$ has
dimensions  of  inverse  time  and is  the  difussion  coefficient (the
stochastic  process $\sigma(t)$  is dimensionless).   Given  that such
stochastic  process  is  introduced  as  a model  of  quantum  gravity
fluctuations at the  Planck scale we expect that for  both $D$ and $k$
the ``natural''  value is  of the order  of the quantum  gravity scale
$\kappa$.

With this correlator and the  relation (\ref{medcorr}), it is not hard
to see that
\begin{equation}
\label{fincorr}
 \langle \sigma^2 \rangle _c=\frac{D}{ k}\left(e^{-k\Delta}-1\right) \frac{1}
{k^2\Delta^2} +\frac{D}{k}\frac{1}{k\Delta}.
\end{equation}

In   our   model,   we   have  five   stochastic   processes,   namely
$\xi,\eta,{\xi^L}_0$,  each one  characterized by  two  parameters ---
$D_{\xi},k_{\xi},  D_{\eta},k_{\eta},D_{\xi^L\,_0},k_{{\xi^L}_0}$  ---
and only one scale of resolution $\Delta$. Fixing these scales, we can
make contact  with different DSR  proposals.  In this example  we will
consider,  however,  that  the  correlation  time  is  equal  for  all
processes $k_{\xi}=k_{\eta}=k_{{\xi^L}_0}\equiv k$.

Since the resolution scale is the one at which spacetime is probed ---
equivalently,  it is  the scale  at  which the  particle averages  the
gravitational  fluctuations ---  this scale  must be  greater  than the
correlation  time  in  order  to  give  sense to  a  time  average  as
(\ref{titet}).  We  then  assume  $\Delta \gg  k^{-1}$  and  therefore
neglect   the   contribution    coming   from   the   exponential   in
(\ref{fincorr}). In this limit we have
\begin{equation}
\label{fincorrmax}
 \langle \sigma^2 \rangle _{c}\sim\frac{D}{ k}\left(\frac{1}{k\Delta}+\frac
{1}{k^2\Delta^2}\right)\qquad \mbox{for}\quad k\Delta \gg 1.
\end{equation}
Then, the non linear relations (\ref{meden}) and (\ref{medmom})
turn out to be
\begin{eqnarray}
E &\sim&{\cal E}\left( 1-\frac{{\cal D}}{2k^2}\frac{1}{\Delta}+\cdots
\right),
\label{enex}
\\
{\bf p}&\sim&{\boldsymbol \pi}\left( 1-\frac{D_\eta}{2k^2}\frac{1}{
\Delta}+\cdots
\right),
\label{momex}
\end{eqnarray}
where ${\cal D}=D_\xi-\sum_L D_{{\xi^L}_0}$ and $\sum_L D_{{\xi^L}_0}$
is   the   sum  of   the   diffusion   coefficient  corresponding   to
${\xi^L}_0$.

{From} these expressions we see that, for DSR1, the choice ${\cal D}/k^2
=1/\kappa$ and $D_\eta =2{\cal  D}$, is mandatory, while ${\cal D}/k^2=
1/\kappa$ and $D_\eta  = {\cal D}$ reproduce DSR2  expressions.
{  Note that  the above  requirements  are fully  consistent with  our
expectation that both $D$ and $k$ are constants of order $\kappa$.  In
both cases we are assuming that the scale of resolution $\Delta$ is of
the order of  the inverse of the true energy of  the particle, that is
$\Delta^{-1}\sim  {\cal E}$.  As  commented above  this choice  of the
scale of  resolution appears to introduce a  frame dependence. Clearly
only  a full treatment  of QG  can tell  whether this  is the  case or
not. What can  be said here is that, if the  correlators above are the
same in all reference frames,  no frame dependence will be introduced.
Notice  also that this  is fully  consistent with  the identifications
above with the invariant scale.

As we said, in order to give sense to the average, we have to consider
that the time during which the particle probes the spacetime is large
compared with the correlation time. Still, it is interesting to
consider the limit case when $\Delta$ is of the order of $k^{-1}$.
The effective  correlator is  of order ${\cal  D} /k $  when $k\Delta$
tends  to 1.  This means  that the  corrective term  to  the effective
tetrad (metric) is of the same order of the flat one. Therefore the QG
effects mimicked by the  stochastic terms we introduced make spacetime
totally undetermined. In this sense  $k^{-1}$ assumes the meaning of a
minimal length.

\subsection{White noise}

Another interesting (albeit unphysical)  example of fluctuation is the
{\em  white noise}. This  can be  obtained from  the previous  case by
asking  that  in the  limit  $k\to  0$  ${\cal D}/k^2  =\gamma$,  with
$\gamma$ a  finite constant with dimensions of a time. In  this case the
correlators turn out to be Dirac delta functions
\begin{equation}
\label{wn}
 \langle \sigma(s)\sigma(s') \rangle ={\gamma}\delta(s-s').
\end{equation}

All integrals are trivial and it is straightforward to check that the non
linear functions for measured energy and momentum are given by
\begin{eqnarray}
E&=&{\cal E}\left(1-\frac{1}{2\Delta}(\gamma_\xi
-\gamma_s)\cdots\right),
\\
P&=&{\boldsymbol \pi}\left(1-\frac{\gamma_\eta}{2\Delta}\cdots\right),
\end{eqnarray}
with $\gamma_s=\sum_L\gamma_{{\xi^L}_0}$.
Again, a formal identification $\gamma_\xi -\gamma_s=\kappa^{-1}$ and
$2(\gamma_\xi -\gamma_s)=\gamma_\eta$ gives DSR1 and
$\gamma_\xi -\gamma_s=2\kappa^{-1}$ with
$(\gamma_\xi -\gamma_s)=\gamma_\eta$
correspond to DSR2 case.

All our previous arguments are
applicable here, and the identification of $\gamma$ as the invariant
scale has the same physical origin as in the previous case.

In synthesis, the examples considered  here give rise, at first order,
to DSR type non linear  relations between the measured energy momentum
and the classical energy momentum of the particle.

\section{Conclusions}
 
In this note we discussed a specific, although schematic, 
realization of the ideas expressed in \cite{libson, libnoi}, namely
that the emergence of modifications to Special Relativity is related
to the measurement and in particular to the average that any particles
operates on the Quantum Gravitational fluctuations of spacetime.  

In our model,  a stochastic term is   added to the tetrad
of a  flat space time and two  averages are defined. One  is the usual
statistical average, defined through  the distribution function of the
stochastic  processes.  The  other   takes  into  account  the  finite
resolution that a particle propagating on this spacetime can probe.

These  modifications  give  rise  to  an effective  metric,  once  all
averages are performed, with terms that depend on the resolution scale
of  the   particle,  much  like  the  {\it   rainbow  gravity  models}
\cite{rainbow}.  This dependence  appears only through the correlation
functions  of the  stochastic  processes, since  we  assumed that  the
dynamical variables are  the tetrads rather then the  metric.  A first
comment is interesting here, namely  that the particle propagates on a
flat  background {\em  only}  if the  correlators,  which contain  the
resolution scale,  identically vanish. This  gives a condition  on our
(simplified) model for the {\it exact} validity of Special Relativity.

The effective metrics allows to define the corresponding tetrad and
then,  non linear  relations  between measured  momenta and  classical
variables  easily follow. Up to this point our approach has been
completely general, modulo the simplifying assumption of Gaussian
behaviour of the stochastic process.
In particular, the condition for a frame independent situation {\it \`a
  la} DSR, contrasted with straight violation of LI, corresponds to the
frame independence of the correlators. It is at the level of the
correlators that a new scale, which we assume to be connected to the
Planck one, emerges. 

In order  to make  contact with DSR  (or LIV)  we need to  specify the
resolution scale $\Delta$. In standard Quantum Mechanics (QM) it would
be proportional to $1/\cal E $.  But it is conceivable that QM will be
modified well before reaching the  QG scale, where the geometry itself
may loose meaning. So, once  $\Delta\sim1/\cal E$ is assumed, we trust
the validity of  our relations only to first  order in the fluctuation
parameters
\footnote{However  once  this assumption  is  made,  it  is no  longer
possible to distinguish DSR from  LIV.}, and given a particular set of
stochastic processes,  we find that  it is possible to  mimic DSR-type
relations (or better their first order limit).

After the parameters  of the correlators are fixed  in order to obtain
the desired  DSR-like behaviour, the  form of the effective  metric is
fixed. For instance,  it is immediate to check that in  the DSR2 it is
diagonal,  giving  as expected  a  light-like  behaviour for  massless
particles.  The  previous results depend on the  identification of the
(would  be)  invariant  scale  with  a  specific  combination  of  the
diffusion coefficient and the correlation time.

We  have presented  here  a concrete  example  of how  the average  on
fluctuations  of  spacetime could  give  rise  to DSR-like  dispersion
relations,  although we  cannot state  what is  the ``right''  DSR (if
there is  one) from  the cases studied  here.  Our examples  should be
taken as an argument in favor of  the idea that DSR might appear as an
effective feature of quantum gravity effects.

\section*{Acknowledgements}
This  work  was partially  supported  by  grant  1060079 from  Fondecyt
(Chile) (F.M.).

\end{document}